\documentclass[sigconf, nonacm]{acmart}
\AtBeginDocument{%
  }

\usepackage{listings}%
\usepackage{colortbl}
\usepackage{subcaption}
\usepackage[most]{tcolorbox}

\usepackage{orcidlink}

\tcbset{
	mycode/.style={
		colback=white,
		colframe=gray!50,
		boxrule=0.2mm,
		arc=2mm,
		left=2mm,
		right=2mm,
		top=0mm,
		bottom=0mm,
		listing only,
		enhanced,
	}
}

\begin{document}

\title[Model-Driven Pipeline for Data Quality Specification and Operationalization]
{A Model-Driven Pipeline for Data Quality Specification and Operationalization: A No-Code Approach for Domain Experts}

\author{Arno Kesper}
\affiliation{%
	\orcidlink{0000-0002-5042-1087}
	\orcid{https://orcid.org/0000-0002-5042-1087}
	\institution{Verbundzentrale des GBV, Göttingen}
	\institution{Philipps-Universität Marburg}
	\city{Marburg}
	\country{Germany}}
\email{Arno.Kesper@gbv.de}

\author{Lukas Sebastian Hofmann}
\affiliation{%
	\orcidlink{0009-0004-3823-3894}
	\orcid{https://orcid.org/0009-0004-3823-3894}
	\institution{Philipps-Universität Marburg}
	\institution{Universidad Complutense de Madrid}
	\city{Marburg}
	\country{Germany}}
\email{Lukas.Hofmann@uni-marburg.de}

\author{Markus Matoni}
\affiliation{%
	\orcidlink{0000-0003-4389-5871}
	\orcid{https://orcid.org/0000-0003-4389-5871}
	\institution{Verbundzentrale des GBV, Göttingen}
	\city{Göttingen}
	\country{Germany}
}
\email{Markus.Matoni@gbv.de}

\author{Gabriele Taentzer}
\affiliation{%
	\orcidlink{0000-0002-3975-5238}
	\orcid{https://orcid.org/0000-0002-3975-5238}
	\institution{Philipps-Universität Marburg}
	\city{Marburg}
	\country{Germany}
}
\email{Taentzer@mathematik.uni-marburg.de}

\renewcommand{\shortauthors}{Kesper et al.}

\newcommand{\class}[1]{\texttt{#1}}

\renewcommand*{\figureautorefname}{Fig.}
\newcommand{\lstlistingautorefname}{Lst.}
\renewcommand*{\sectionautorefname}{Sec.}
\let\subsectionautorefname\sectionautorefname
\let\subsubsectionautorefname\sectionautorefname


\begin{abstract}

High-quality data is essential for reliable analysis, decision-making, and research across domains.
This is especially relevant in areas such as cultural heritage, where data is collected and curated manually, making it prone to quality issues like inconsistencies.
To improve data quality, the data must be analyzed regularly using systematic quality analyses.
Quality analyses validate the conformance of data to domain-specific expectations.
These expectations are best understood by domain experts, who can express them using natural language.
However, they rarely possess the technical expertise to formalize these expectations into executable quality analyses.  
Consequently, this process requires domain experts and data engineers, making it time-consuming and technically demanding.
The required technical expertise and the resulting dependencies pose a significant challenge.

To address this challenge, we present a pipeline for formalizing and operationalizing data quality constraints.
We support this pipeline using QPM, a metamodel for defining templates for reusable quality analyses.
The web application Constrainify enables tailoring templates to specific conceptual requirements and translating them into executable quality analyses via a tool-chain based on model-driven engineering subpipelines.
The result is a set of reusable, repeatable, and domain-specific quality analyses.

\end{abstract}

\maketitle

\section{Introduction}

Data quality is essential for effective use of domain-specific data.
In many fields of application, such as cultural heritage, healthcare, and life science large datasets are collected, curated, and maintained by alternating stakeholders over long periods of time.
Data quality issues, such as inconsistencies, incompleteness, or wrong statements, severely impact data quality, which is understood as 'fitness for use'~\cite{wang1996}.
The detection of data quality issues is essential for ensuring and maintaining reliable and usable data~\cite{pipino2002}.

However, the definition of quality analysis in domain-specific settings presents a significant challenge.
Data quality problems are inherently domain-specific \cite{strong1997}, so there is no out-of-the-box solution for quality analyses.
Instead, individual problem identification and formalization is based on domain-specific expertise.
It is helpful to understand how these problems manifest in the data, which requires a basic knowledge regarding the data model and representation.
Transferring this knowledge into quality analyses requires technical expertise in specific database technologies, data processing infrastructures, query languages, and schema languages \cite{kesper2020}.
Common database technologies include relational databases, XML~\cite{xml}, RDF~\cite{rdf}, and Neo4j\footnote{Neo4j \url{https://neo4j.com/} (2026-07-02)} data.
These technologies come with specialized query and schema languages, such as
SQL~\cite{sql} for relational databases, XQuery\footnote{XML Query Language (XQuery) \url{https://www.w3.org/TR/xquery/} (2026-07-02)}, XSD\footnote{XML Schema Definition Language (XSD) \url{https://www.w3.org/TR/xmlschema11-1/} (2026-07-02)}, and Schematron\footnote{Schematron \url{https://www.schematron.com/} (2026-07-02)} for XML, SPARQL\footnote{\label{note_sparql}SPARQL Query Language \url{https://www.w3.org/TR/sparql11-query/} (2026-07-02)} and SHACL\footnote{Shapes Constraint Language (SHACL) \url{https://www.w3.org/TR/shacl/} (2026-07-02)} for RDF, and Cypher\footnote{Neo4j Cypher \url{https://neo4j.com/docs/cypher-manual} (2026-07-02)} for Neo4j.

The required expertise for this process is typically distributed across multiple roles, specifically domain experts and data engineers. %
Domain experts possess a solid understanding regarding semantics and quality requirements of data, but often lack the technical skills to implement quality analyses~\cite{pham2010}.
In contrast, data engineers can implement quality analyses, but rarely possess the domain knowledge for identifying quality issues.
Domain experts always delegate quality analyses to data engineers.
Consequently, the process of defining, implementing and conducting data quality analysis requires close coordination among multiple stakeholders \cite{collaborative}.
These dependencies result in additional effort, reduce efficiency and hinder dynamic adaptation of changing requirements.

We conclude that there is a need for approaches that reduce the dependencies between the involved roles and lower the amount of technical knowledge required for specifying and executing data quality analyses.
The approaches should enable domain experts to contribute their knowledge more directly while minimizing the need for technical skills.
Further, they should decouple the execution of quality analyses from data processing infrastructures.

To cope with this challenge, we present a tool-supported pipeline for the formalization and execution of domain-specific data quality analyses.
This novel pipeline translates domain knowledge into executable quality checks while minimizing the coordination effort between domain experts and data engineers.

To support the pipeline, we present a novel use case of the Quality Pattern Model (QPM) \cite{kesper2020} metamodel for templates that specify quality analysis independent from database technologies.
These templates can be adapted automatically to different database technologies, such as XML, RDF, and Neo4j.
In the adaptation process, parameters are introduced to specify domain-specific requirements.
If all parameters are specified, an artifact can be operationalized for repeatable quality analysis.
Constrainify \cite{constrainify_whitepaper}, a web application based on QPM, supports the different steps of the pipeline.

We present the following contributions:

\begin{enumerate}
	\item A 4-step-\textbf{pipeline} for an interactive specification of executable quality analyses based on identified quality issues.
	\item Tool-support for the pipeline using the \textbf{Quality Pattern Model} (QPM) metamodel \cite{kesper2020} and the web application Constrainify \cite{constrainify_whitepaper}.
	\item An end-to-end pipeline \textbf{evaluation} through a case study.
\end{enumerate}

This paper is structured as follows:
We motivate our work in \autoref{sec_quality-problems} and explain our 4-step-pipeline for data quality analyses in \autoref{sec_pipeline}.
The pipeline depends on templates, based on the QPM metamodel, presented in \autoref{sec_qpm}.
An interactive web application is presented in \autoref{sec_constrainify}.
\autoref{sec_evaluation} presents a case-based evaluation.
Related work is discussed in \autoref{sec_related-work}. 
Finally, \autoref{sec_conclusion} concludes our paper.
\section{Data Quality Problems}
\label{sec_quality-problems}

Data quality is typically defined as 'fitness for use'~\cite{wang1996}.
Thus, quality is a spectrum depending on how well it suits the intended purposes.
It is typically described through a set of \textit{quality dimensions}~\cite{qualitysurvey, qualitydimensions, wang1996, wand1996, strong1997, batini2009, cai2015}.
Commonly discussed dimensions include completeness, accuracy and consistency.
The dimensions capture different characteristics that contribute to overall usability.
However, the interpretation of the dimensions into concrete requirements depends on the domain.
Consequently, data quality requirements can vary between domains, organizations, and use cases.
As such, an extensive data quality assessment must capture aspects from multiple perspectives.

\subsection{Quality Problems in Cultural Heritage Data}

The domain of cultural heritage is an example, where data is often collected manually and integrated from several institutions to be published in online portals.
Examples are the German Digital Library \footnote{\label{ddb}``Deutsche Digitale Bibiliothek'' (German Digital Library, DDB) \url{https://www.deutsche-digitale-bibliothek.de/?lang=en} (2026-07-02)} (DDB) or the Marburg Bildindex \footnote{``Bildindex der Kunst \& Architektur'' (Picture Index of Art and Architecture) \url{https://www.bildindex.de/} (2026-07-02)} of the German Documentation Center for Cultural Heritage\footnote{German Documentation Center for Art History (DDK) \url{https://www.uni-marburg.de/de/fotomarburg} (2026-07-02)} (DDK).
For the data to be integrated, the portals need to analyze the data for a base quality.

Based on six qualitative interviews and a workshop with 19 domain experts who work in the acquisition, management and usage of cultural heritage in different organizations, Kesper et al. \cite{zenodo-catalog} compiled a comprehensive catalog of data quality issue types.
Each problem type is systematically documented with possible causes, identification, and its impact on data quality.
The catalog encompasses 12 groups of data quality problems with several subgroups.
They include missing and incorrect data, redundancies, and violations of the data format.

An analysis of this catalog reveals, that many quality problems follow recurring patterns.
Substantial methodical overlap can be observed among the problem identifications across some groups.

\subsection{Definition of Quality Analysis}

We see the definition of data quality analysis as a multi-step process.

First, relevant quality requirements must be \textit{identified}.
A data quality requirement specifies a quality property, that data is expected to satisfy to be of high quality \cite{iso-dataquality, wang1996}.
Requirements can be derived from various sources, including specific data quality issues, the intended use of data, and stakeholder expectations.
To generalize a problem into a requirement, an understanding of the domain and the data semantics is required.
\\
Second, the requirements must be translated into a plan \textit{how} to conduct a quality analysis.
This includes determining how a requirement violation can be identified within the data structure via one or more data quality constraints, which are formal, measurable rules, that operationalize a requirement.
This requires an understanding of the underlying database technology and the data format.
\\
Third, the analysis plan must be \textit{implemented} as executable analysis.
This implementation can be done using a query language capable of executing the specified analysis plan against the data.
For example, XML data can be analyzed using the language XQuery.
This step requires technical knowledge regarding the specific query language.
\\
Finally, the specified analysis can be \textit{executed} on a target dataset.
The detected violations can be interpreted and used to continuously improve the data.

The different steps for defining a quality analysis require some specialized expertise, namely domain-specific understanding and technical skills.
Generally, this expertise is not combined within a single person.
Instead, the identification of a data quality issue and its generalization into a requirement are tasks for a domain expert.
The operationalization of requirements is a task for a data engineer.
To enable the data engineer to do so, the domain expert has to discuss the requirement with a data engineer, explaining the conditions and occurrences in detail.
Miscommunication can result in multiple iterations.
Traditionally data engineers execute the quality analysis, commissioned by domain experts.

Whenever a new requirement arises, a new analysis needs to be defined.
When the new requirement exhibits the logic of an existing analysis, the data engineer may be able to take over some implementation, saving a little bit of work.
This requires the data engineer to keep an overview over the defined analyses.
Every time that an analysis needs to be repeated, the data engineer needs to be commissioned anew.

This process requires a lot of consultations and coordination.
The effort could be reduced when enabling domain experts to define, manage, and execute analyses themselves, without needing extensive technical knowledge.

\subsection{Running Example}
\label{subsec_running-example}

In the following, we present one specific quality requirement as running example, which targets the reference validity of specific values in LIDO\footnote{\label{lido}``Lightweight Information Describing Objects'' (LIDO) \url{https://icom-documentation.mini.icom.museum/working-groups/lido/lido-overview/about-lido/what-is-lido/} (2026-07-02)} data.
LIDO is an XML format that emerged in the context of digital humanities for delivering metadata across organizations.

\vspace{-1,5mm}
\begin{tcolorbox}[mycode]
\vspace{-0,2cm}
\begin{lstlisting}[label=lst_example_requirement,caption={Quality requirement example}, breaklines=true]
In LIDO data, every Link Resource Element must contain a resolvable URL.
\end{lstlisting}
\vspace{-0,2cm}
\end{tcolorbox}
\vspace{-1,5mm}

To analyze this requirement on LIDO data, we need to iterate over all link resource values in the data and validate the contained values.
If a stated URL cannot be resolved, this instance is a quality issue.
Such quality issues need to be identified and reported to domain experts to undergo quality improvement.
The same requirement also applies to different elements in LIDO, such as \lstinline|Legal Body Identifier| and \lstinline|Record Info Links|.
A possible template to express how the quality analysis works is:

\vspace{-1,5mm}
\begin{tcolorbox}[mycode]
Search for <properties> that are not a valid link.
\end{tcolorbox}
\vspace{-1,5mm}

This formulation describes logic for violation detection.
We found that domain experts prefer constraints to be formulated positively.
A fitting constraint reformulation is: 

\vspace{-1,5mm}
\begin{tcolorbox}[mycode]
Each <property> must be a valid link.
\end{tcolorbox}
\vspace{-1,5mm}

This running example presented here is used throughout the paper to illustrate the proposed operationalization pipeline.

\section{Pipeline}
\label{sec_pipeline}

In this section, we propose a model-driven pipeline, that operationalizes natural language templates into repeatable executable quality analyses on domain-specific data.
The tool-based pipeline provides a structured process for transforming domain-specific data quality knowledge into formal requirements and iteratively define a repeatable data quality analysis via executable constraints.
To manage the complexity of this transformation and to separate concerns, the pipeline is decomposed into four interconnected but autonomous subpipelines:
Requirement Elicitation, Constraint Specification, Template Creation and Quality Analysis.

The pipeline starts with \emph{Requirement Elicitation}, where domain experts identify quality problems in databases, and formulate semi-formal requirements.
If a fitting template exists, domain experts can continue with the \emph{Constraint Specification} by operationalizing the requirements into constraints in an agile manner.
If there is no fitting template for the requirement, a data engineer needs to be define a new one via the \emph{Template Creation} pipeline.
Specified constraints enable \emph{Quality Analyses} that can be executed repeatedly by data analysts for iterative quality improvements.

\begin{figure}
	\centering
	\includegraphics[width=\linewidth]{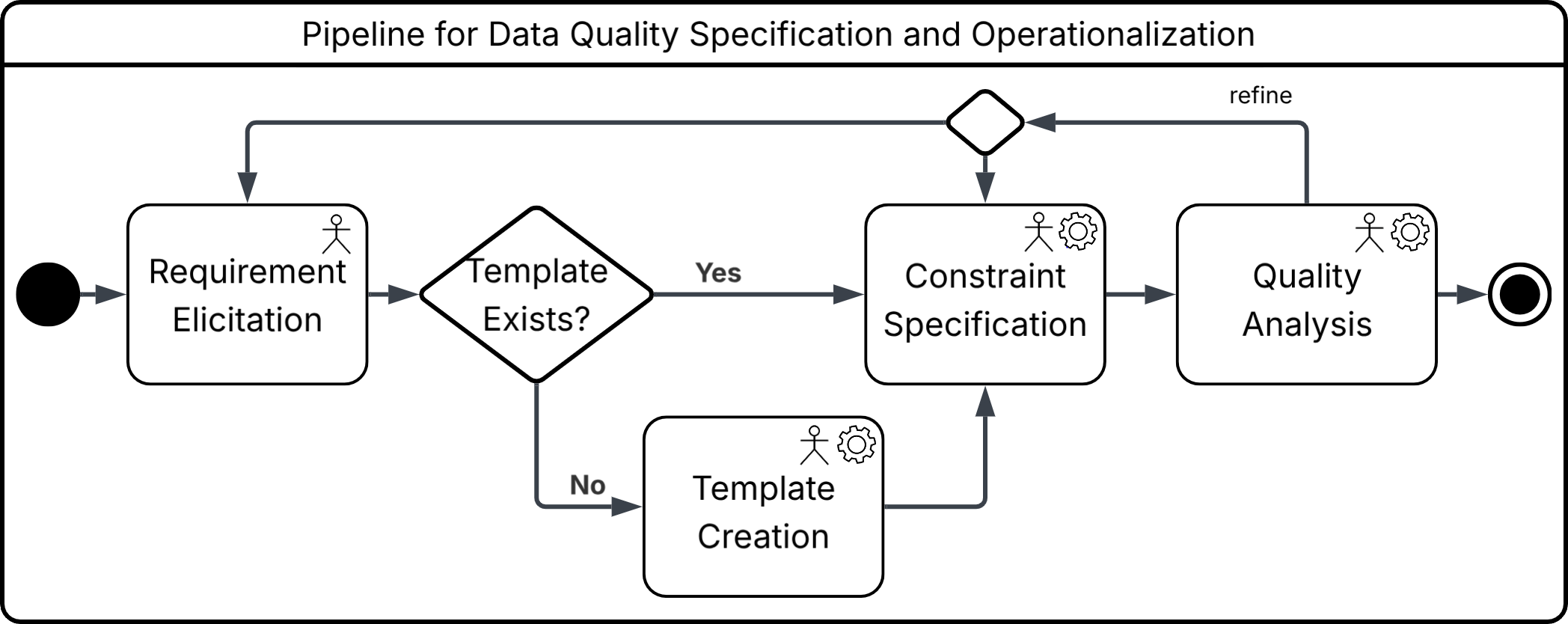}
	\vspace{-0.6cm}
	\caption{Data Quality Specification and Operationalization}
	\label{fig_pipeline}
	\vspace{-0.4cm}
\end{figure}

\autoref{fig_pipeline} illustrates, how these stages support the systematic translation from informal quality requirements to executable data quality constraints. %
Some subpipelines depend on specific artifacts created by another subpipeline, but they do not need to be executed continuously, as existing artifacts can be reused in different contexts. %

\subsection{Roles}
\label{pipeline_roles}

The pipeline includes four user roles, namely domain experts, power users, data engineers, and data analysts.

\textit{Domain experts} are people who host, manage, and curate one or more databases.
Their work focuses on collecting and integrating data for the database and improving its quality.
They have a comprehensive overview of the data, and extensive informal knowledge regarding its quality criteria.
With this point of view, they are able to identify inconsistencies across the data set.
However, they are not necessarily familiar with technology, databases, or data schemas.

\textit{Power users} actively work with the data mainly as information resource.
For example, this includes researchers, that analyze the data to discover new knowledge.
They dig deep into the provided data and are the first to identify small quality problems.

\textit{Data engineers} offer analytical abilities and technical skills.
They are able to abstract from a problem and formalize it.
This role also includes the assessment of capabilities of existing technologies.
Their main task in this pipeline is to grasp the requirements and formalize them into quality analyses.

\emph{Data Analysts} comprise users who execute data quality analyses.
Traditionally, analyses were executed by Data Engineers, as it was necessary to operate specific data processing infrastructures.
As our tool-supported pipeline reduces the technical barriers, it is expected to decouple the role Data Analysts from Data Engineers.
Based on our tool support, we anticipate Domain Experts to be able to execute predefined quality analyses themselves.

\subsection{Requirement Elicitation}
\label{pipeline_requirements}

\begin{figure}
	\centering
	\includegraphics[width=\linewidth]{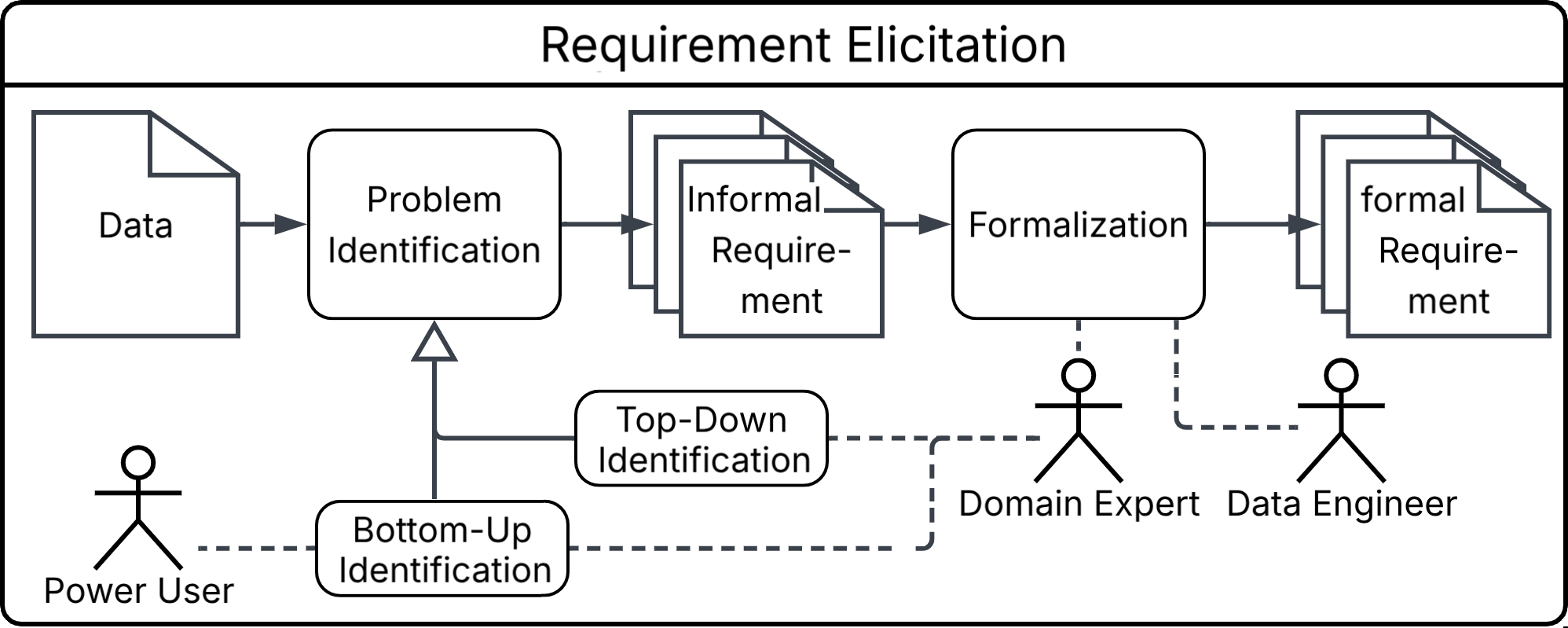}
	\vspace{-0.6cm}
	\caption{Requirement Elicitation Pipeline (Manual)}
	\label{fig_pipeline-requirement}
	\vspace{-0.3cm}
\end{figure}

The \textit{Requirement Elicitation} pipeline, as illustrated in \autoref{fig_pipeline-requirement}, is a manual task that delineates a structural exploration and identification of requirements for specific databases.

The process is initiated with the \textit{problem identification} in the databases.
The identification of requirements in a database is often a passive process of gaining informal knowledge through various methods.
The primary method of specifying requirements is Top-Down identification, during the data planning stages, prior to database creation.
Domain experts, who collect, integrate, and manage data in databases, are familiar with the domain-specific expectations and are able to define optimal data from a bird's-eye perspective.
\\
Commonly, problems in existing databases are identified reactively through a bottom-up approach during the use and analysis of the data.
Domain experts and power users that intensively work with the data frequently identify minor problems.
In certain cases, these problems go deeper and are caused by underlying issues.
Knowledge about specific data quality issues qualifies as an informal data quality requirement.

Next, this informal requirement (e.g., \autoref{lst_example_requirement}) needs to be formalized.
Domain experts need to identify a way to formulate it explicitly in a verifiable manner as a constraint.
Further it should be checked for technical feasibility, optionally with a data engineer.
The outcome of this pipeline is a formal requirement.

\subsection{Constraint Specification}
\label{pipeline_constraint}

\begin{figure}
	\centering
	\includegraphics[width=\linewidth]{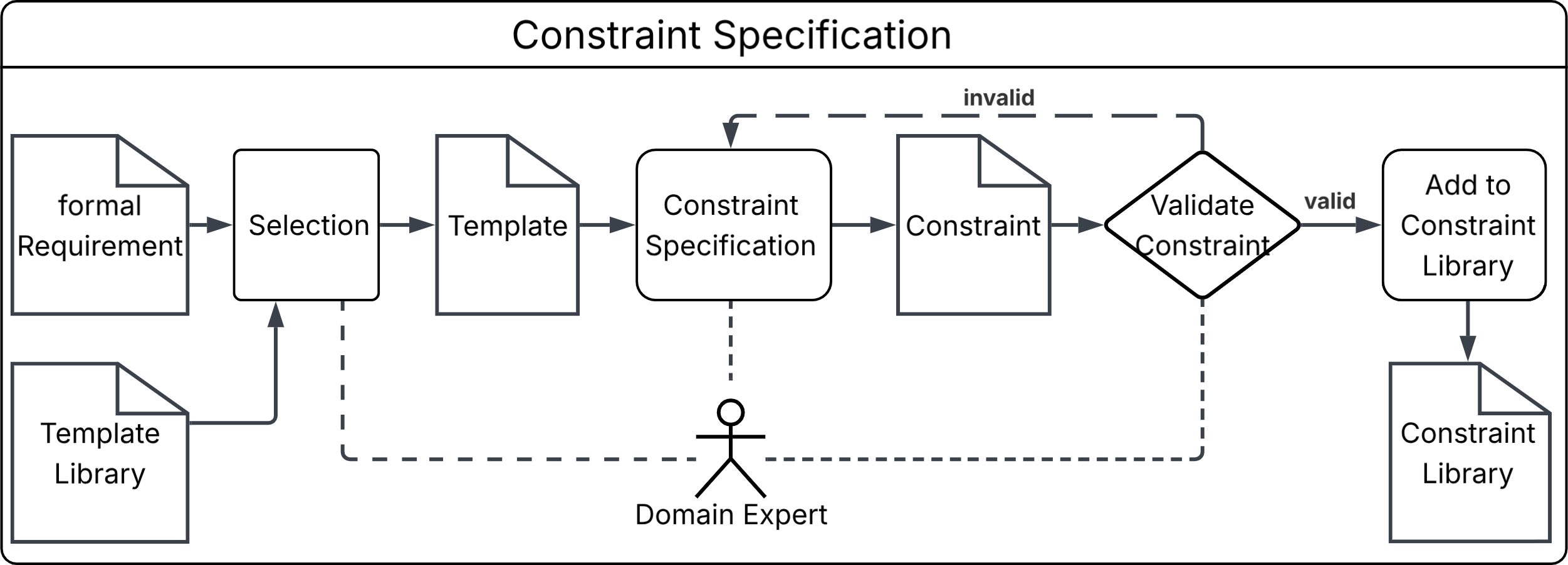}
	\vspace{-0.6cm}
	\caption{Constraint Specification Pipeline (Semi-Automated)}
	\label{fig_pipeline-constraint}
	\vspace{-0.5cm}
\end{figure}

In this semi-automatic subpipeline, formal data quality requirements are used to specify quality analyses as constraints based on templates.
\autoref{fig_pipeline-constraint} illustrates the process for the specification of constraints using a template library.
Each template defines an abstract quality analysis for a group of quality problems.
One example for a template is shown in \autoref{subsec_running-example}.
Templates can be adapted to each problem instance in the class.

Starting with a formal requirement (e.g., \autoref{lst_example_requirement}), a domain expert selects the template from a preexisting template library that best matches the intended analysis logic.
Next, a constraint can be specified from the template by specifying a set of parameters to cover the requirement.
A constraint is a fully adapted template, which includes the logic for a specific quality analysis.

Whenever a new constraint is created, it needs to be validated.
Primarily this includes testing, if it does what it is supposed to do.
Further, it also includes duplicate detection.
If the constraint is validated, it can be added to a constraint library, which defines a set of executable quality analyses.
To enable easy overview and selection, the constraints can be annotated with tags or sorted into constraint sets.

\subsection{Template Creation}
\label{pipeline_template}

\begin{figure}
	\centering
	\includegraphics[width=\linewidth]{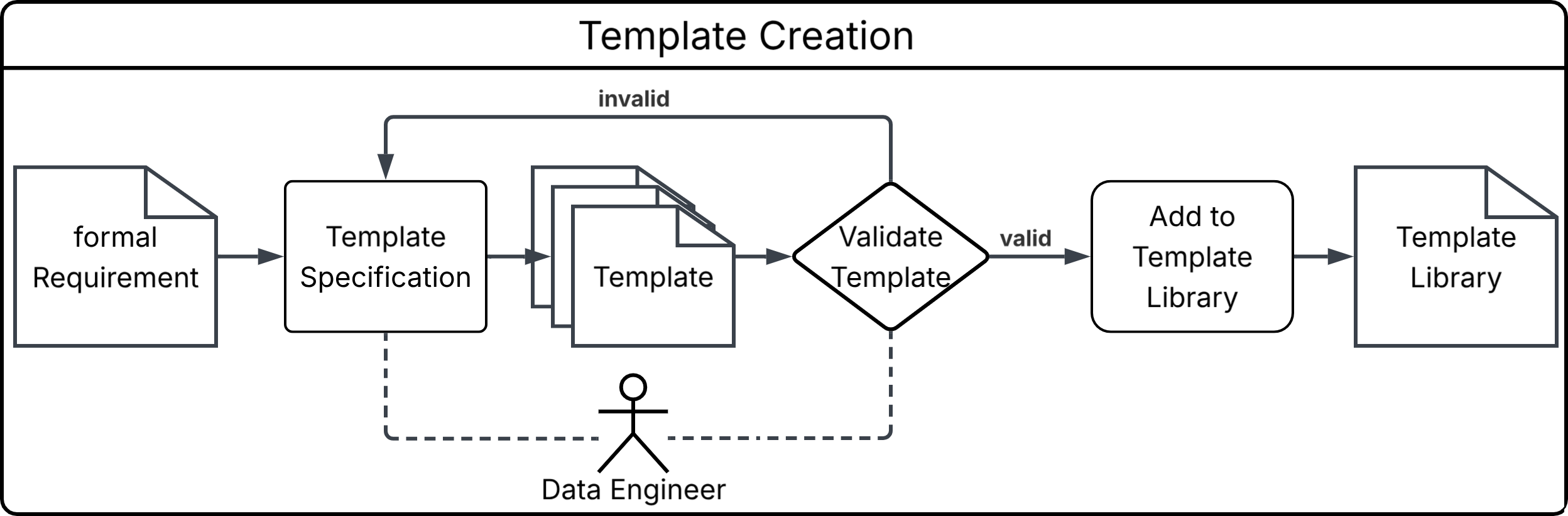}
	\vspace{-0.6cm}
	\caption{Template Creation Pipeline (Semi-Automated)}
	\label{fig_pipeline-template}
	\vspace{-0.4cm}
\end{figure}

If there is no fitting template for a specific formal requirement, a new template needs to be created, as illustrated in the \emph{Template Creation} pipeline in \autoref{fig_pipeline-template}.

Creating a quality analysis template from a formal requirement is a manual task that requires technical abilities, thus needs to be handled by a data engineer.
Given the requirement as a natural language sentence, the first step is to identify how violations of the requirement can be detected using a search query.

A newly created template must encode the search query using logical constructs and operators.
These constructs and operators contain parameters, which must not be specified in the parameterizable template.
Certain parameters can be predefined if necessary or useful.
Finally, a representation of the template is defined as a constraint sentence with gaps for parameters (c.f., \autoref{subsec_running-example}).
Although the template is represented as a constraint sentence, it encapsulates the logic to detect constraint violations.

A newly created template needs to be validated, by verifying its functionality and making sure, it is not a duplicate to an existing template.
If the check succeeds, the new template can be stored in the template library to be available for domain experts.
If the check fails, the template needs to be revised.

\subsection{Quality Analysis}
\label{pipeline_analysis}

\begin{figure}
	\centering
	\includegraphics[width=\linewidth]{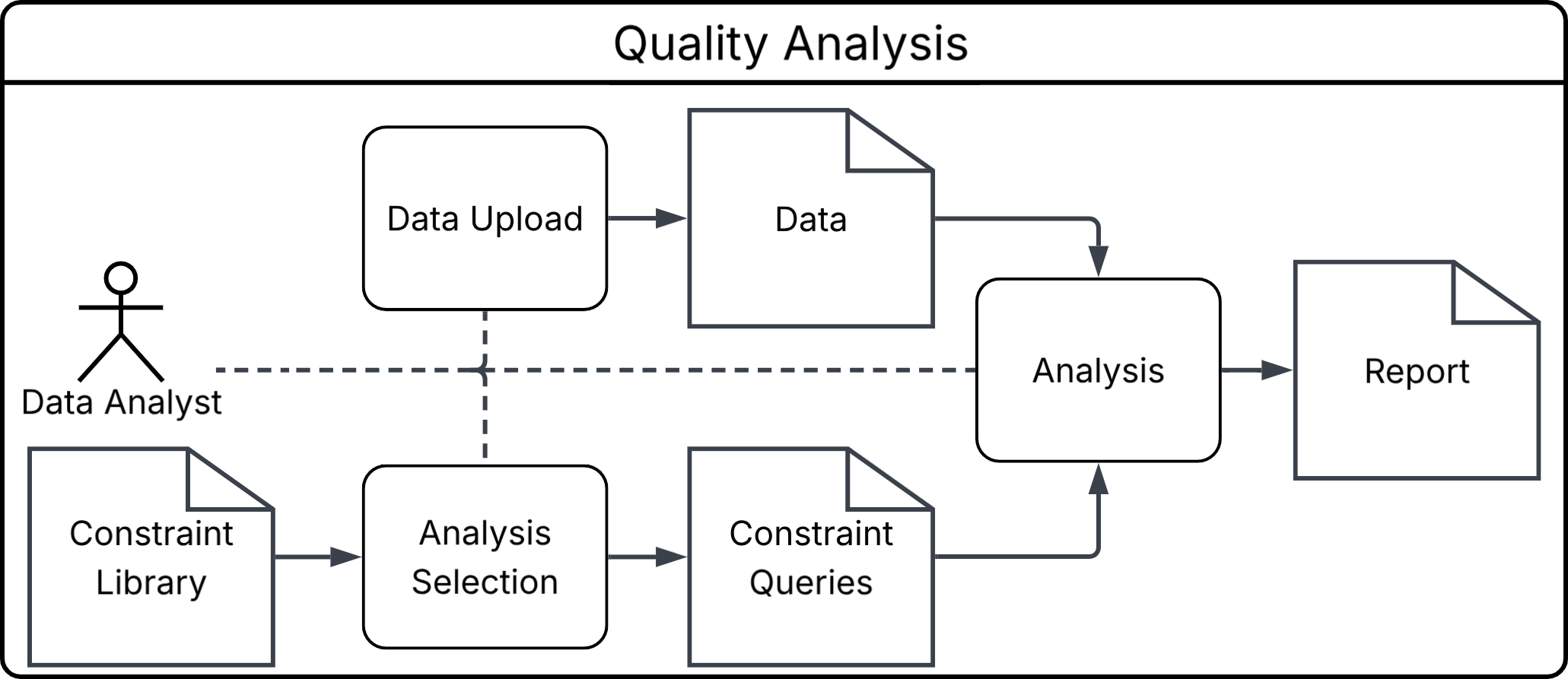}
	\vspace{-0.6cm}
	\caption{Quality Analysis Pipeline (Semi-Automated)}
	\label{fig_pipeline-analysis}
	\vspace{-0.4cm}
\end{figure}

All constraints defined according to \autoref{pipeline_constraint} are collected in a constraint library.
Each constraint defines a repeatable quality analysis.
The semi-automated subpipeline for operationalizing a quality analysis is presented in \autoref{fig_pipeline-analysis}.

The data quality analysis is executed by a data analyst.
Based on the constraint library, the data analyst can select a data analysis.
This selection process  can be more easy using provided tags or constraint sets.
From a given constraint set, the analyses can be created as a set of constraint queries.
Secondly, the data to be analyzed needs to be specified.

Given a selected constraint set and data, the quality analysis can be started.
Hereby, the data is evaluated against every selected constraint and a detailed analysis report is created.
This report can be seen as instructions for quality improvement.

\section{Quality Pattern Model}
\label{sec_qpm}

\begin{figure*}
	\centering
	\includegraphics[width=\linewidth]{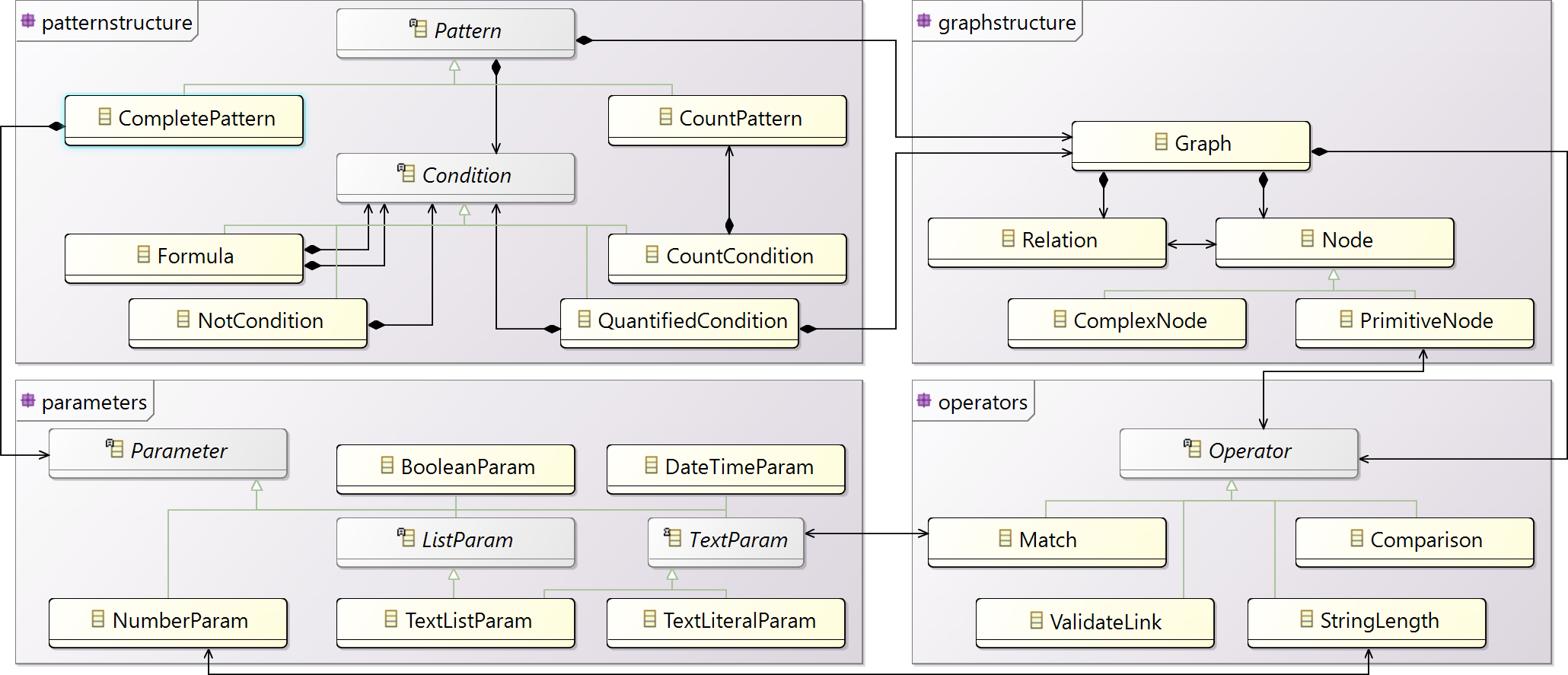}
	\vspace{-6mm}
	\caption{Condensed Class Diagram for QPM}
	\label{fig_class-diagram}
	\vspace{-2mm}
\end{figure*}

The proposed pipeline for operationalizing data quality requirements into executable quality analyses utilizes templates and constraints.
This section introduces the metamodel for such templates and constraints.
The model-driven approach reduces the need for technical expertise in database technologies and query languages.

The model-driven approach \emph{Quality Pattern Model} (QPM) addresses this challenge \cite{kesper2020}.
The implementation is available on GitHub\footnote{Quality Pattern Model (QPM) Repository \url{https://github.com/Project-KONDA/pattern-based-quality-analysis} (2026-07-02)} and Zenodo \cite{zenodo-qpm}.

\subsection{Concepts}

QPM enables the definition of generic query templates that encapsulate the complete analysis logic while remaining independent of database technology.
These QPM templates hold the logic for anti-patterns to analyze data by identifying requirement violations.
Thereby, QPM is able to reduce the required knowledge regarding technologies and query languages. 
This also enables the application of templates to different technologies.
\\
The adaptation of generic templates to different database technologies is done fully automatic.
The implementation currently comprises XML, RDF, and Neo4j data.
Furthermore, an adaptation for relational databases is already mapped out.
Result of the adaptation is an abstract template.

Abstract templates come with a small set of parameters, allowing users to tailor the template to a specific quality problem in a specific database. %
To specify a constraint for a domain-specific requirement, a value needs to be assigned to each parameter.
Based on a brief template documentation, even domain experts without technical knowledge should be able to specify constraints.

A fully specified QPM constraint can be translated into an executable database query fitting to the database technology.
This database query is ready to be applied to selected data in order to identify requirement violations.

\subsection{QPM Metamodel}
\label{subsec_metamodel}

The QPM metamodel enables the definition of pattern-based quality analyses as machine-interpretable artifacts.
Furthermore, the metamodel provides the functionality to adapt, specify, and translate the templates into executable quality analyses.
A condensed version of the QPM metamodel is shown as a class diagram in \autoref{fig_class-diagram}.
According to Kleppe \cite{dsl}, this metamodel can be seen as a domain-specific language (DSL) for defining templates, as it defines an abstract and a concrete syntax and semantics.

The core class of the QPM metamodel is \class{Complete\-Pattern}, which defines the logic for a generic QPM template.
The \class{Complete\-Pattern} holds a structure of \class{Conditions} over \class{Graph}s.
\class{Condition}s are defined using a Composite Pattern.
\class{Graph}s contain \class{Node}s and \class{Relation}s, which are mapped to data elements of different technologies in later stages.
Hereby, structural elements of a database technology are represented as \class{Complex\-Node}s, while single values are represented as \class{Primitive\-Node}s.
\class{Node}s can be annotated using \class{Operator}s, specifying restrictions onto data values.
In this state, the \class{Complete\-Pattern} is generic, meaning it defines a quality analysis independent from any database technology.

When adapting a specified \class{Complete\-Pattern} to a specific database technology, the pattern is supplemented with technology-specific parameters.
Using the parameters, the mapping from the graphs to specific data is determined.
For example, XML uses XPath parameters for addressing the relationships between related nodes.
In this state, the \class{Complete\-Pattern} specifies an abstract template, now specific to a technology.

If all parameters of an abstract template are specified with a validated value, the \class{Complete\-Pattern} represents a concrete constraint.
This constraint is ready to be applied to data.

\subsection{Modeling the Running Example}
\label{qpm_running-example}

In this section, we demonstrate, how to express the requirement from \autoref{subsec_running-example} using the QPM metamodel.

The general query logic involves searching for values in the data, that fulfill the condition of not representing a resolvable link.
The query logic, can be seen as independent from any database technology, as no technology-specific particularities are mentioned.

\begin{figure*}[thbp]
	\small
	\centering
	\begin{subfigure}[t]{.31\linewidth}
		\centering
		\includegraphics[width=.9\linewidth]{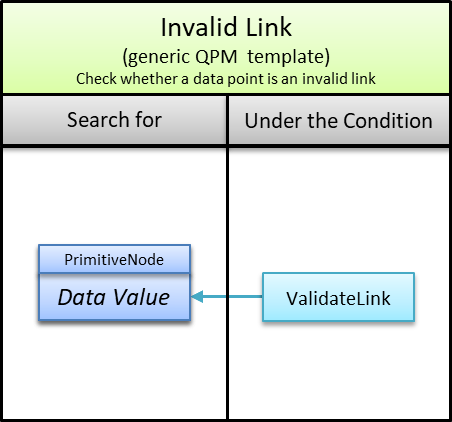}
		\caption{Generic QPM template}
		\label{fig_link_generic}
	\end{subfigure}
	\begin{minipage}[t]{.035\linewidth}
	\end{minipage}
	\begin{subfigure}[t]{.31\linewidth}
		\centering
		\includegraphics[width=.9\linewidth]{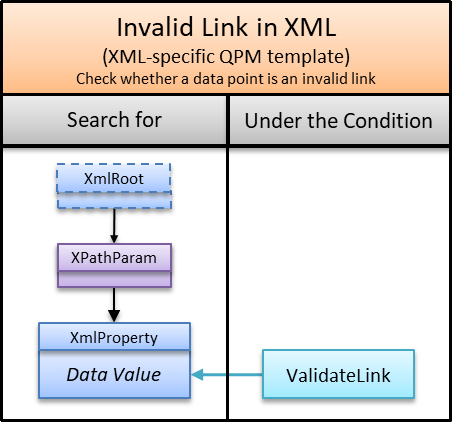}
		\caption{XML-specific QPM template}
		\label{fig_link_xmlabstract}
	\end{subfigure}
	\begin{minipage}[t]{.035\linewidth}
	\end{minipage}
	\begin{subfigure}[t]{.31\linewidth}
		\centering
		\includegraphics[width=.9\linewidth]{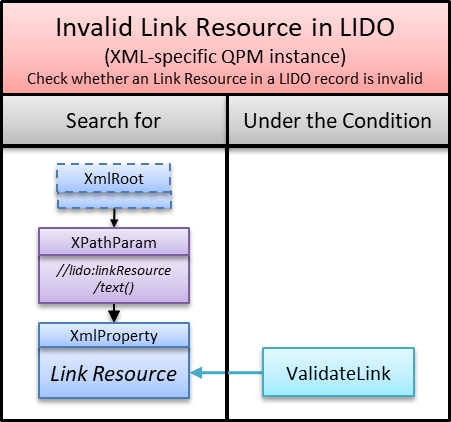}
		\caption{LIDO-specific QPM instance}
		\label{fig_link_xmlconcrete}
	\end{subfigure}
	\vspace{-0.3cm}
	\caption{QPM models for \mbox{identifying} invalid links, realizing the running example from \autoref{subsec_running-example}}
	\vspace{-2mm}
\end{figure*}

The QPM metamodel allows us to implement the generic logic in a model-driven manner.
Here, we specify, that we are searching for specific data values, represented by a \class{Primitive\-Node}.
This \class{PrimitiveNode} is annotated with the operator, that the value must represent an invalid link, thus is not resolvable.
The resulting generic QPM template is an object structure visualized in \autoref{fig_link_generic}.

To apply this template, we start by adapting it to a specific database technology.
For our example, we choose XML.
This adaptation process is done fully automatically.
In case of XML, the adaptation process transforms the \class{Node}s into XML-specific equivalents.
Further, it adds an \class{XmlRoot} element and adds a \class{Relation} to be able to address the specified value.
Finally, it adds a parameter to specify an XPath for the \class{Relation}.
The result is an XML-specific (abstract) QPM template, see \autoref{fig_link_xmlabstract}.

This abstract template is left with only one parameter, specifically the \class{XPathParam} for specifying which kind of data values are meant via  an XPath.
With respect to our running example, we can specify the parameter with the LIDO\footref{lido}-specific XPath to the LIDO Link Resource.
The result is a concrete QPM instance, as visualized in \autoref{fig_link_xmlconcrete}.
A QPM instance can be translated automatically into an executable artifact, which is applicable to any LIDO data.

We represent QPM templates as natural language sentences in which we embed the parameters.
This way the parameters can be considered in the context of the query logic.
Our running example (s. \autoref{qpm_running-example}) matches exactly the following natural sentence shown in \autoref{subsec_running-example}.
Thus, the QPM instance from \autoref{fig_link_xmlconcrete} can be represented using the sentence:

\vspace{-0,2cm}
\begin{tcolorbox}[mycode]
Each \fcolorbox{gray}{white}{Property} must be a valid link. \\
\small Property [XPath to XMLProperty]: \textit{//lido:linkResource/text()}
\end{tcolorbox}
\vspace{-0,2cm}

To be able to apply this template to XML data, this sentence must be parameterized with an XPath that targets an XML Property.
The resulting executable artifact analyzes the data for all violations, thus detects invalid links, and produces a fitting quality report.

\subsection{QPM Framework}

To enable practical application, an implementation of the metamodel is integrated into a lightweight framework and distributed as a Docker image with a REST API.
It can be used to build an integrated tool chain for the pipeline that supports managing QPM templates and instances as a model-driven engineering pipeline.
It enables specification of parameters for templates and the translation of specified QPM instances to executable artifacts.
Finally, the artifacts can be applied to data for conducting a data quality analysis.

\subsection{Discussion}

The QPM metamodel establishes a foundation for a template-based approach to data quality analysis.
It builds the structure for a framework, which can split the subpipelines for creating analyses into defining templates and tailoring the templates to concrete requirements.
The implemented functionality of the metamodel takes over query generation, resulting in the fact that neither of the new subpipelines require knowledge about formulating queries.

Templates can be defined generically, i.e., independently of any database technology, and implemented and tested independently of specific requirements.
Given that templates can cover multiple quality analyses, it reduces the overall effort of data engineers.
Using a set of predefined templates \cite{zenodo-templateset} can further reduce the work.

When defining specific constraints, a template is selected and the parameters are filled.
The constraint definition only has technical dependencies in the case where templates are missing.
Given a set of constraints, a repeatable analysis is defined, which can be handled uniformly.
This gives the opportunity to improve consistency, traceability, and maintainability of quality analyses.

In summary, QPM provides the possibility to define templates as machine-interpretable artifacts.
When fully specified, the artifacts can be transformed into executable quality analyses.
QPM aims to eliminate a lot of technical knowledge required and needs for consultation and improve the overall quality of specified quality analyses.
Thus, QPM specifically supports the subpipeline for \emph{Template Specification} (see \autoref{pipeline_template}) and lays the foundation for the subpipelines \emph{Constraint Specification} and the \emph{Quality Analyses}.
\section{Constrainify}
\label{sec_constrainify}

With the QPM metamodel we have the basis for defining templates and executing quality analysis.
However, a framework with a REST-API is not usable by domain experts, as it still relies on technical abilities.
Thus, the framework for QPM was extended with the web application Constrainify, to enable domain experts to work with the templates.
The code is published on GitLab\footnote{Constrainify Repository \url{https://gitlab.gwdg.de/aqinda/constrainify} (2026-07-02)} and Zenodo \cite{zenodo-constrainify}.
Additionally, we provide a Demo-Instance\footnote{Demo-Instance: 'Constrainify - Simplify creating quality constraints and analyzing your data' \url{https://constrainify.gwdg.de/} (2026-07-02)}.
This tool is supposed to support domain experts with a no-code formalization and application of requirements.
Specifically, this application offers tool support for the pipeline steps for \textit{Constraint Specification} (cf. \autoref{pipeline_constraint}) and \textit{Quality Analysis} (cf. \autoref{pipeline_analysis}):
Based on templates provided by QPM and a conceptual understanding of requirements, domain experts shall be enabled to define and execute data analyses.

\subsection{Constraint Specification}

To derive a constraint from a requirement, the user is guided through a constraint-specifying process.
The process begins with an overview of the predefined templates from QPM using the provided representation as natural language sentences with one or more gaps, that correspond to the parameters of the template.
These are sentences in the domain-specific language
for the structured representation and formalization of data quality knowledge.
To facilitate navigation, these sentences are grouped according to their quality dimension, such as Completeness, Cardinality and Reference Validity.
Additionally, the parameter gaps may be prefilled with example values to illustrate the intended semantics.
Users select a template sentence that best matches their intended requirement from the overview.

\begin{figure}
	\centering
	\includegraphics[width=\linewidth]{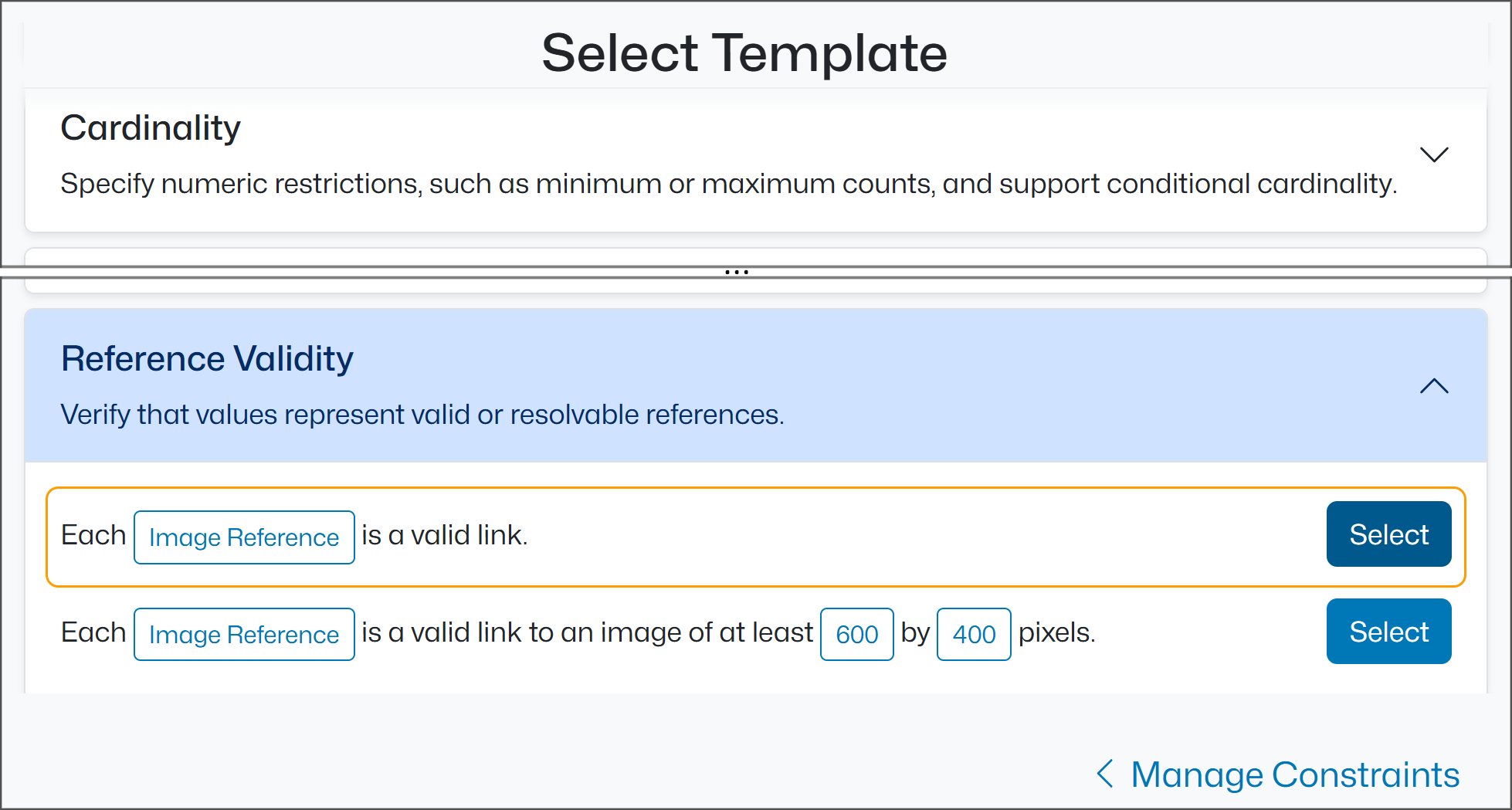}
	\vspace{-0.6cm}
	\caption{Template Selection in Constrainify}
	\label{fig_constrainify-templates}
	\vspace{-0.3cm}
\end{figure}

An excerpt from the template selection is presented in \autoref{fig_constrainify-templates}.
In the running example (see \autoref{subsec_running-example}), we want to analyze whether the XML element ``Link Resource'' in a LIDO\footref{lido} data record contains a valid URL.
Thus, the group \emph{Reference Validity} is considered, where a template is identified that corresponds to the specified requirement.

After selecting a template, the user is directed to a page where the template can be fully instantiated.
Here, the constraint can be assigned a name, and values can be provided for all template parameters.
The application offers parameter-specific assistance depending on the parameter type.
As a result, users are only required to provide high-level input, while the technical implementation of the constraint is handled by the application.
Consequently, the application constitutes a no-code approach for the operationalization of data quality requirements.
For example, when an XML schema is available, XML-related parameters can be identified from arbitrary textual inputs using an NLP-based mapping approach.
Once all parameters have been specified, the template is instantiated, resulting in a concrete data quality constraint, which can be stored in the Constraint library.

\begin{figure}
	\centering
	\includegraphics[width=\linewidth]{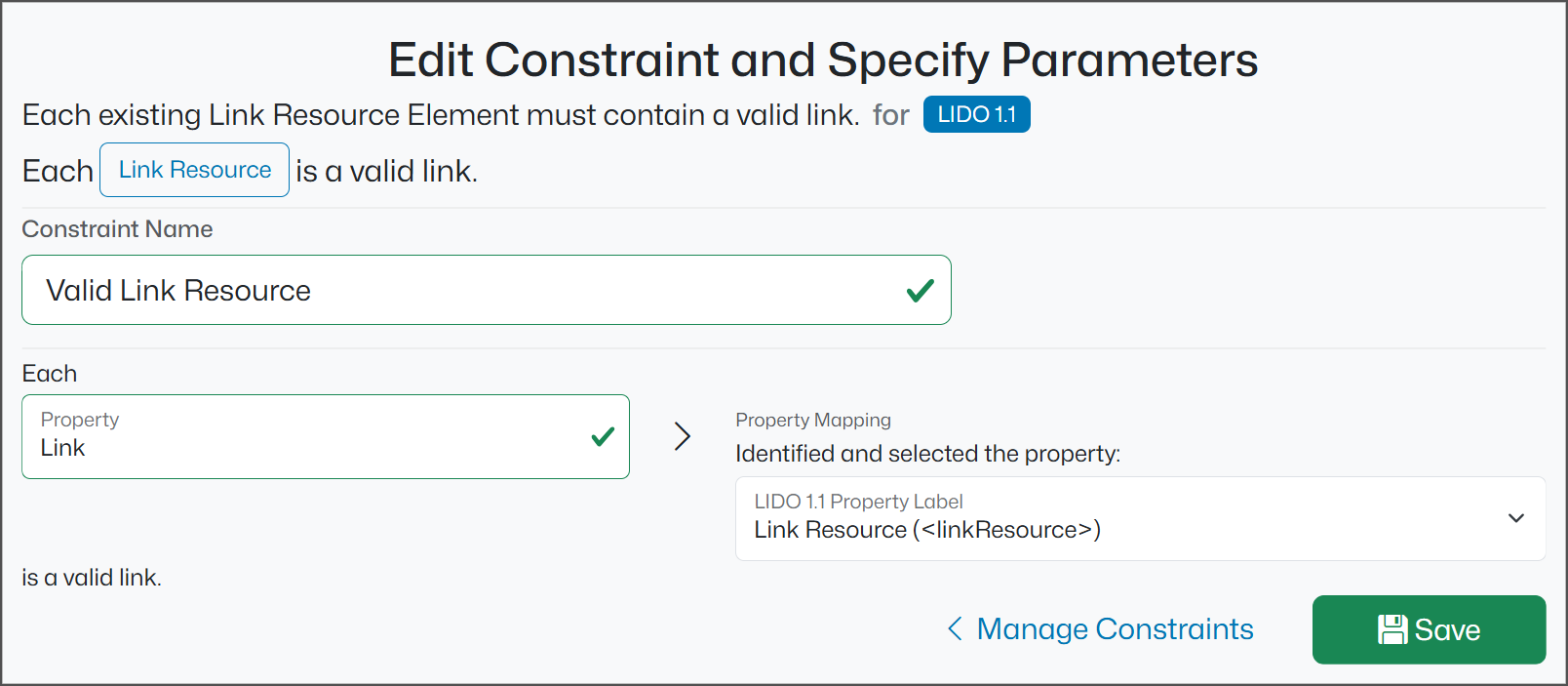}
	\vspace{-0.6cm}
	\caption{Constraint Specification in Constrainify}
	\label{fig_constrainify-constraint}
	\vspace{-0.5cm}
\end{figure}

The instantiation process is presented in \autoref{fig_constrainify-constraint}.
In the case of the running example, we name the Constraint according to our requirement and specify our parameter.
In this case, it is sufficient to enter the term 'Link' to identify the correct XML-element called Link Resource.
The result is a constraint that fits to our requirement.

\subsection{Quality Analysis}

After a constraint has been specified, it becomes available in the constraint library.
Each constraint can be operationalized as an executable artifact that enables repeatable data quality analyses.
To perform an analysis, users upload a data set to be assessed and select one or more constraints from the constraint library, that fit to the data format.
The uploaded data set is then assessed against the selected constraints, resulting in a quality report.
The report summarizes the evaluation results and presents the detected quality issues.
For each detected issue, the report contains snippets of the data, line numbers, and additional information on the problem.
This enables users to systematically assess data quality and repeatedly apply the same constraint set to different data sets and versions.

\begin{figure}
	\centering
	\includegraphics[width=\linewidth]{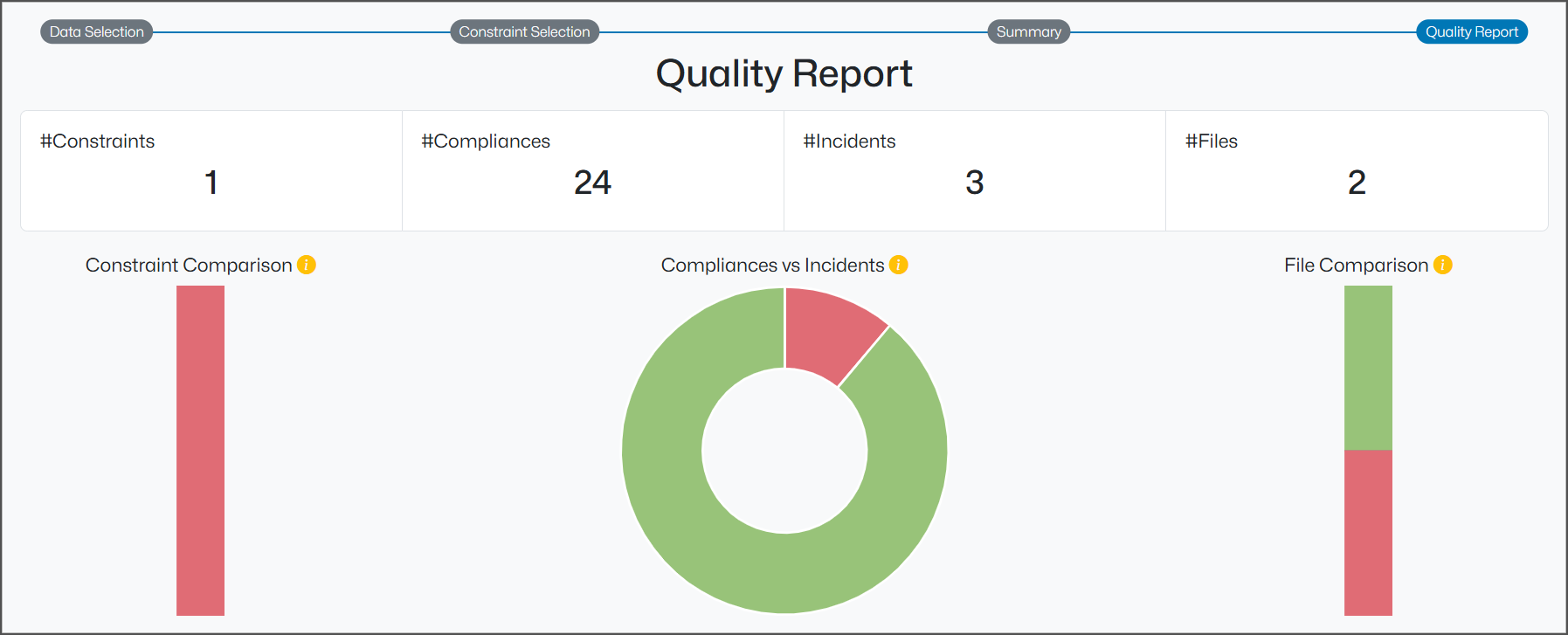}
	\vspace{-0.6cm}
	\caption{Quality Report in Constrainify}
	\label{fig_constrainify-quality-report}
	\vspace{-0.4cm}
\end{figure}

In our running example, we analyze two LIDO\footref{lido}-files with our defined constraint.
After a quality analysis, a report is generated, as shown in \autoref{fig_constrainify-quality-report}.
In our case, we were able to identify 24 Link Resource elements that contain valid links, but also 3 links that were unable to be resolved.
Also, all incidents only occurred in one of the two files.

\section{Evaluation}
\label{sec_evaluation}

The goal of our pipeline is to enable the specification and operationalization of data quality analyses from natural language requirements.
We complement this process with QPM and the web application Constrainify as specialized tool support.

\subsection{Design}

The German Digital Library (DDB)\footref{ddb} collects data from different organizations, which include different media libraries, cultural heritage preservation organizations and museums.
In their day-to-day work, they work extensively with data to import and get to know the requirements for their data very well.
In the process, they documented requirements for their delivery data\footnote{Requirements \url{https://deutsche-digitale-bibliothek.atlassian.net/wiki/spaces/DFD/pages/48103977/Anforderungen+an+die+Lieferdaten} (2026-07-02)}.
This gives us a representative set of requirement statements from practice.
That document is independent of the structure of the data only targeting the information delivered.

For the evaluation, we aim to formalize this set of requirements for the LIDO-XML\footref{lido} data format by walking through our pipeline (cf. \autoref{sec_pipeline}) starting with zero templates.

\subsection{Research Questions}
\label{questions}

We formulated the following research questions, each corresponding to one subpipeline:

\begin{enumerate}
	\item [\textbf{RQ1:}] To what extent can implicit data quality requirements be formalized as explicit and verifiable requirements?
	\item [\textbf{RQ2:}] To what extent can reusable templates be derived from formally specified data quality requirements?
	\item [\textbf{RQ3:}] To what extent can executable constraints be generated through template instantiation?
	\item [\textbf{RQ4:}] To what extent can a constraint library be operationalized for comprehensive and repeatable data quality analysis?
	\item [\textbf{RQ5:}] To what extent is the developed tool support usable by domain experts?
\end{enumerate}

\subsection{Execution}

To validate the proposed tool-supported pipeline (\autoref{sec_pipeline}), we perform a case-based validation using the practical requirements document for delivery data from ABC as the primary input artifact.
As the document specifies the expectations for data quality independently of any data format, it represents a realistic starting point for the proposed requirement-driven process.
The goal of the case-based pipeline validation is the creation of an executable constraint set for the LIDO\footref{lido} data format.

\paragraph{\textbf{Requirement Elicitation}}

The document comprises a total of 55 individual requirement statements in natural language.
Out of these, we identified 10 statements as out of scope.
Hereby, 6 statements were withdrawn during discussions, 2 statements were not testable, as they targeted the immutability of values.
Finally, 2 statements were not applicable for the LIDO\footref{lido} data format.
The remaining 45 requirement statements were subjected to the formalization step (cf. \autoref{pipeline_requirements}).

In this step, we identified 18 (40\%) out of the 45 requirement statements as formulated at a high level of abstraction, i.e., encompassing multiple verifiable conditions.
Some requirements each can be decomposed into two to six constraints, resulting in 29 additional ones.
This process resulted in $74$ ($45 + 29$) formal requirements.

\begin{tcolorbox}[mycode]
	\paragraph*{\textbf{Answer for RQ1}}
	Most implicit data quality requirements can be systematically formalized as explicit, verifiable, and machine-processable requirements.
	In the evaluation, 82\% of the requirements were successfully formalized, while 40\% required decomposition into multiple requirements.
\end{tcolorbox}

\paragraph{\textbf{Template Creation}}

In this subsequent process, we iterated through each formalized requirement and tried to instantiate a fitting template (cf. \autoref{pipeline_constraint}) via Constrainify (\autoref{sec_constrainify}).
As we started with zero templates, we were required to implement a template library from scratch.
Note that the template definition and the constraint specification were carried out by distinct project members.

Given the 74 formal requirements, we grouped them into 19 distinct template types, with one to 12 requirements per group, averaging $3.9$ constraints per template.
Of these template types, 16 were implemented and integrated into our Constrainify instance. %

The first template targets uniqueness of identifiers, which requires an analysis across the whole data set, which can contain multiple files.
The current implementation only supports uniqueness checks within the same file, thus we can only analyze this constraint to a certain degree.
The remaining two templates require non-deterministic natural language processing.
One requires detection of whether a text is written in a specific natural language, the other describes an evaluation on whether a textual description targets a specific artifact or only a more generic category.

The results indicate, that a relatively small set of generic templates is sufficient to formalize and operationalize the vast majority of the data quality requirements formulated in the document of the DDB.

\begin{tcolorbox}[mycode]
	\paragraph*{\textbf{Answer for RQ2}}
	Reusable templates can be systematically derived by abstracting recurring structural patterns from formally specified data quality requirements. 
	The evaluation identified 19 reusable templates, of which 84\% were implemented.
\end{tcolorbox}

\paragraph{\textbf{Constraint Definition}}

The implemented templates were sufficient to realize 69 of the 74 formal requirements as constraints, corresponding to a recall of 93.2\%.
The remaining 5 requirements are associated with the three currently unsupported templates.
Thus, unsupported requirements stem from limitations of the current implementation rather than from limitations of the pipeline or the template-based approach.
All 69 constraints were correct, resulting in an F1-score of 96.5\%.

\begin{tcolorbox}[mycode]
	\paragraph*{\textbf{Answer for RQ3}}
	Executable constraints can be generated through template instantiation with high accuracy.
	The implemented template set reached an F1-score of 96.5\% with an average of 3.9 constraint per template. %
\end{tcolorbox}

\paragraph{\textbf{Quality Analysis}}

We were able to apply the 69 constraints to real, received datasets, provided by the DDB.
These datasets encompassed 25 files.

The 69 constraints did analyze a total of 2,668 findings, i.e., instances matching the test condition in the data.
Each constraint did verify between 11 and 76 matching instances.
38 constraints did not detect any quality issues, while the remaining 31 constraints did identify a total of 245 quality issues.

\begin{tcolorbox}[mycode]
	\paragraph*{\textbf{Answer for RQ4}}
	A constraint library can be operationalized to perform repeatable and automated data quality analysis. %
	In the evaluation, it detected 245 quality issues across 25 real-world data files and produced structured quality reports.	
\end{tcolorbox}

\paragraph{\textbf{User Study}}
To evaluate usability, we conducted a small qualitative user study with scenario-based exercises \cite{zenodo-referencedata}.
We tasked eight domain experts to define two specific constraints and perform the corresponding quality analyses, followed up with a structured questionnaire \cite{zenodo-questionaire} focused on usability.
The participants completed the tasks only with minor hesitations and praised the intuitiveness and clarity of the UI.

\begin{tcolorbox}[mycode]
	\paragraph*{\textbf{Answer for RQ5}}
	The developed tool support is usable by domain experts.
	All participants successfully specified correct constraints and executed the corresponding quality analyses, indicating that the tool can be used effectively after a short introduction.
\end{tcolorbox}

\subsection{Threats to Validity}

This case-based pipeline evaluation is an initial feasibility study.
Threats to \textit{internal validity} arise from subjective interpretation of intermediate results, which may introduce researcher bias.
\textit{External validity} is limited because the evaluation is based on a single requirements document from the cultural heritage domain.
Although this document is representative of its domain, the results cannot be generalized without further empirical evaluations using additional requirement documents from different domains.
\section{Related Work}
\label{sec_related-work}

Existing work on data quality management addresses different stages of the process, ranging from requirement specifications to executable data quality analyses.
However, these stages are typically supported by isolated approaches rather than integrated workflows.
This section reviews work on natural language template specification and data quality assessment frameworks and positions the proposed pipeline.

\subsection{Natural Language Template Specification}

The Wikidata Query Service\footnote{Wikidata Query Service \url{https://query.wikidata.org} (2026-07-02)} allows the definition of natural language query templates for Wikidata by combining SPARQL queries with sentence templates sharing common variables.
The Query Helper renders these templates as parameterized natural language sentences with selectable values.

Similarly, BigCQ \cite{BigCQ} provides a database of reusable natural language templates for competency questions that are mapped to predefined SPARQL queries.
FRASE \cite{frase} instead relies on LLMs to translate unrestricted natural language into SPARQL queries for Wikidata.
While this enhances flexibility, the authors report limited accuracy (maximum F1 score across all experiments of 50 \%) and a strong dependence on high-quality textual descriptions of the relations and classes.

Although these approaches target semantic query generation rather than data quality, they demonstrate that parameterized natural language templates can serve as an effective abstraction layer between domain experts and formal query languages.
Their application to the operationalization of data quality requirements, however has not yet been investigated.
Further, they are only applicable to specific database technologies.
Approaches supported by LLM also introduce a significant risk of unreliability.

\subsection{Data Quality Assessment}

MQAF \cite{mqaf1} and SodaCore\footnote{SodaCore \url{https://github.com/sodadata/soda-core} (2026-07-02)} emphasize declarative rule specification through YAML-based configuration formats,
whereas Great Expectations\footnote{Great Expectations \url{https://greatexpectations.io/} (2026-07-02)} and Deequ\footnote{Deequ \url{https://github.com/awslabs/deequ} (2026-07-02)} \cite{deequ-ml, deequ} primarily define executable data quality tests in code.
Deequ further supports the automatic derivation of data constraints through data profiling.
More recently, Soda\footnote{Soda \url{https://soda.io/} (2026-07-02)} and Great Expectations have introduced AI-assisted support for generating validation rules from metadata and natural-language requirements.
In general, LLM extensions for data quality assessment seem promising, but not yet fully developed \cite{frameworks-llmevaluation}.
While these developments indicate the potential of LLM-supported data quality assessment, current solutions primarily assist rule authoring and still require manual translation of data quality requirements into executable quality constraints.

\subsection{Summary}

Existing approaches typically support either the specification of natural language-based templates or the execution of data quality analyses.
However, the transformation from data quality requirements into executable data analyses remains largely manual, while LLM approaches have not yet achieved sufficient reliability.
As a result, the effort for quality analysis definition is increased, while the reuse and traceability of constraints remain limited.

The proposed pipeline (\autoref{sec_pipeline}) addresses this gap by operationalizing data quality requirements through parameterized natural language templates that automatically generate repeatable data quality analyses within an integrated, tool-supported workflow, thereby increasing traceability and reducing manual effort.
\section{Conclusion}
\label{sec_conclusion}

This paper presents a tool-supported pipeline for the iterative specification of repeatable data quality analyses.
The proposed approach structures the transformation of informal data quality requirements into executable quality constraints.

The pipeline is divided into four interconnected subpipelines:
Requirement Elicitation, Template Creation, Constraint Specification, and Quality Analysis.
Separating the pipeline into distinct stages reduces the coordination effort and interdependencies between domain experts and data engineers.
The dedicated tool support is used to reduce the overall technical expertise required and maintain traceability of the intermediate artifacts.

The evaluation demonstrated the feasibility of the proposed pipeline.
Informal data quality requirements could be formalized and abstracted into reusable templates.
The resulting templates could be instantiated into executable constraints that support repeatable data quality analysis.
The specified constraints were applicable to real cultural heritage delivery data.
The results indicate, that a comparatively small set of generic templates can be sufficient to realize the majority of analyzed requirements, enabling efficient constraint specification and reuse.
The current tool support enables the realization of 93.2\% of the analyzed requirements.

Overall, the presented pipeline combines a structured methodology with dedicated tool support to systematically guide the transformation of domain-specific knowledge into automated and repeatable data quality analyses.

In future work, we plan to extend our implementation to support a comprehensive template library and mature the prototype into a production-ready tool.
These efforts are backed by letters of intent from several institutions for a follow-up project.
We also plan to evaluate the approach on additional data formats (e.g., TEI\footnote{Text Encoding Initiative \url{https://tei-c.org/} (2026-07-02)}) and across different domains (e.g., biodiversity).
We further plan to explore how AI can be integrated to assist domain experts within the constraint specification subpipeline.
\section*{Acknowledgment}

This paper was written as part of AQinDa\footnote{``Agile Qualitätssicherung von Metadaten zu kulturellen Objekten im Kontext von Datenintegrationsprozessen'' (AQinDa, \texttt{2023-2025})}, funded by the German Research Foundation (DFG) which aims to develop a continuous data quality management process.
The authors would like to thank Jakob Voß~\orcidlink{0000-0002-7613-4123}, Philipp Wieber~\orcidlink{0009-0006-1321-195X}, and Regine Stein~\orcidlink{0000-0003-3406-5104} for their valuable support and proofreading.
The publication was funded by the Verbundzentrale des GBV (VZG), Göttingen, Germany.

\bibliographystyle{ACM-Reference-Format}
\bibliography{bibliography}

@inproceedings{kesper2020,
	title     = {Detecting Quality Problems in Research Data: A Model-Driven Approach},
	author    = {Kesper, Arno and Wenz, Viola and Taentzer, Gabriele},
	year      = {2020},
	booktitle = {Proceedings of the 23rd ACM/IEEE International Conference on Model Driven Engineering Languages and Systems},
	location  = {Virtual Event, Canada},
	publisher = {Association for Computing Machinery},
	address   = {New York, NY, USA},
	series    = {MODELS '20},
	pages     = {354–364},
	doi       = {10.1145/3365438.3410987},
	isbn      = {9781450370196},
	url       = {https://doi.org/10.1145/3365438.3410987},
	numpages  = {11}
}

@misc{constrainify_whitepaper,
	title     = {Constrainify: Web Application to Specify and Analyze Data Quality},
	author    = {Matoni, Markus and Kesper, Arno and Hoffmann, Lukas and Király, Péter and Schäfer, Domenic and Voß, Jakob and Taentzer, Gabriele},
	month     = mar,
	year      = 2026,
	publisher = {Zenodo},
	organization = {Zenodo},
	doi       = {10.5281/zenodo.18861823},
	url       = {https://doi.org/10.5281/zenodo.18861823},
}

@software{zenodo-qpm,
	author    = {Kesper, Arno and Wenz, Viola and Hofmann, Lukas Sebastian and Voß, Jakob},
	title     = {Project-KONDA/pattern-based-quality-analysis: Quality Pattern Model v1.5.0},
	month     = jun,
	year      = 2026,
	publisher = {Zenodo},
	organization = {Zenodo},
	version   = {v1.5.0},
	doi       = {10.5281/zenodo.20643230},
	url       = {https://doi.org/10.5281/zenodo.20643230},
	swhid     = {swh:1:dir:cbd4c9af310325e7f83b8a149cba43e80f63426d;origin=https://doi.org/10.5281/zenodo.7299020;visit=swh:1:snp:e876d8a2d0a86840724acb96d6a7943eee399d28;anchor=swh:1:rel:8198a0884a5316166fca84b4f9f252eecd8c0c7e;path=Project-KONDA-pattern-based-quality-analysis-62cf1ea},
}

@software{zenodo-constrainify,
	author    = {Matoni, Markus and Kesper, Arno},
	title     = {Constrainify: Environment},
	month     = jun,
	year      = 2026,
	publisher = {Zenodo},
	organization = {Zenodo},
	version   = {0.9.0},
	doi       = {10.5281/zenodo.20560540},
	url       = {https://doi.org/10.5281/zenodo.20560540},
	swhid     = {swh:1:dir:edd05f63e1efbe956a75459e3793ceb83acee4dd;origin=https://doi.org/10.5281/zenodo.20560539;visit=swh:1:snp:a19e6dcc6708cef6aff3133845c46e014e174e71;anchor=swh:1:rel:16767647c08169d531539fdf457109a543540016;path=constrainify-v0.9.0},
}

@misc{zenodo-catalog,
	title     = {Catalog of Quality Problems in Data, Data Models and Data Transformations},
	author    = {Arno Kesper and Markus Matoni and Julia Rössel and Gabriele Taentzer and Michelle Weidling and Viola Wenz},
	month     = apr,
	year      = 2023,
	publisher = {Zenodo},
	organization = {Zenodo},
	version   = 2,
	doi       = {10.5281/zenodo.7757293},
	url       = {https://doi.org/10.5281/zenodo.7757293},
}

@dataset{zenodo-templateset,
	title     = {Constrainify: Template Collection},
	author    = {Kesper, Arno},
	month     = jun,
	year      = 2026,
	publisher = {Zenodo},
	organization = {Zenodo},
	version   = {1.0.0},
	doi       = {10.5281/zenodo.20718083},
	url       = {https://doi.org/10.5281/zenodo.20718083},
}

@dataset{zenodo-referencedata,
	author    = {Matoni, Markus},
	title     = {Constrainify User Study: Reference Data},
	month     = may,
	year      = 2026,
	publisher = {Zenodo},
	organization = {Zenodo},
	doi       = {10.5281/zenodo.19856977},
	url       = {https://doi.org/10.5281/zenodo.19856977}
}

@dataset{zenodo-questionaire,
	author    = {Matoni, Markus},
	title     = {Constrainify User Study: Questionnaire},
	month     = may,
	year      = 2026,
	publisher = {Zenodo},
	organization = {Zenodo},
	doi       = {10.5281/zenodo.20400437},
	url       = {https://doi.org/10.5281/zenodo.20400437}
}

@misc{xml,
	author    = {Ronald Bourret},
	title     = {{XML} and Databases},
	year      = {1999},
	url       = {http://www.rpbourret.com/xml/XMLAndDatabases.htm}
}

@book{sql,
	title     = {A Guide to {SQL} Standard, 4th Edition},
	author    = {C. J. Date and Hugh Darwen},
	address   = {Reading, MA, USA},
	year      = {1997},
	publisher = {Addison-Wesley},
	isbn      = {0-201-96426-0},
	bibsource = {dblp computer science bibliography, https://dblp.org}
}

@incollection{rdf,
	title     = {Foundations of RDF databases},
	author    = {Arenas, Marcelo and Gutierrez, Claudio and P{\'e}rez, Jorge},
	booktitle = {Reasoning Web International Summer School},
	pages     = {158--204},
	year      = {2009},
	publisher = {Springer},
	address   = {Berlin, Heidelberg}
}

@article{collaborative,
	title     = {Domain experts in the loop: Leveraging generative artificial intelligence for interactive data validation in process mining},
	author    = {Julian Armin Dormehl and Robert Andrews and Wolfgang Kratsch and Maximilian Röglinger and Moe Thandar Wynn and Felix Zetzsche},
	journal   = {Information Systems},
	volume    = {140},
	pages     = {102715},
	year      = {2026},
	issn      = {0306-4379},
	doi       = {https://doi.org/10.1016/j.is.2026.102715},
	url       = {https://www.sciencedirect.com/science/article/pii/S0306437926000293}
}

@book{dsl,
	title     = {Software language engineering: creating domain-specific languages using metamodels},
	author    = {Kleppe, Anneke},
	publisher = {Pearson Education},
	address   = {Boston, MA, USA},
	year      = {2008}
}

@techreport{iso-dataquality,
	author    = {{International Organization for Standardization}},
	title     = {ISO/IEC 25012:2008 -- Software engineering -- Software product Quality Requirements and Evaluation (SQuaRE) -- Data quality model},
	number    = {ISO/IEC 25012:2008},
	year      = {2008},
	url       = {https://www.iso.org/standard/35736.html}
}

@article{BigCQ,
	title     = {BigCQ: A large-scale synthetic dataset of competency question patterns formalized into SPARQL-OWL query templates},
    author={Dawid Wiśniewski and Jędrzej Potoniec and Agnieszka Ławrynowicz},
    journal       = {CoRR},
    volume        = {abs/2105.09574},
    year          = {2021},
    url           = {https://arxiv.org/abs/2105.09574}
}

@article{frase,
	title     = {{FRASE}: Structured Representations for Generalizable {SPARQL} Query Generation},
	author    = {Diallo, Papa Abdou Karim Karou and Zouaq, Amal},
	journal   = {CoRR},
	volume    = {abs/2503.22144},
	doi       = {10.48550/ARXIV.2503.22144},
	year      = {2025}
}

@inproceedings{deequ,
	title     = {Unit testing data with deequ},
	author    = {Schelter, Sebastian and Biessmann, Felix and Lange, Dustin and Rukat, Tammo and Schmidt, Phillipp and Seufert, Stephan and Brunelle, Pierre and Taptunov, Andrey},
	booktitle = {Proceedings of the 2019 International Conference on Management of Data},
	pages     = {1993--1996},
	year      = {2019},
	publisher = {Association for Computing Machinery},
	address   = {New York, NY, USA}
}

@misc{deequ-ml,
	author    = {Sebastian Schelter and Philipp Schmidt and Tammo Rukat and Mario Kiessling and Andrey Taptunov and Felix Biessmann and Dustin Lange},
	title     = {DEEQU - Data quality validation for machine learning pipelines},
	year      = {2018},
	url       = {https://www.amazon.science/publications/deequ-data-quality-validation-for-machine-learning-pipelines}
}

@article{frameworks-llmevaluation,
	title     = {Evaluating Data Quality Tools: Measurement Capabilities and LLM Integration},
	author    = {Rehberger, Tobias and H{\"u}tter, Thomas and Ehrlinger, Lisa and W{\"o}{\ss}, Wolfram},
	journal   = {CoRR},
	volume    = {abs/2604.09163},
	url       = {https://arxiv.org/abs/2604.09163}, 
	year      = {2026}
}

@article{pipino2002,
	title     = {Data Quality Assessment},
	author    = {Pipino, Leo L. and Lee, Yang W. and Wang, Richard Y.},
	year      = 2002,
	month     = apr,
	journal   = {Communications of the ACM},
	volume    = {45},
	number    = {4},
	pages     = {211--218},
	issn      = {0001-0782, 1557-7317},
	doi       = {10.1145/505248.506010}
}

@inproceedings{pham2010,
	title     = {Discovering dynamic integrity rules with a rules-based tool for data quality analyzing},
	author    = {Thanh Thoa Pham Thi and Markus Helfert},
	booktitle = {Proceedings of the 11th International Conference on Computer Systems and Technologies and Workshop for PhD Students in Computing},
	pages     = {89--94},
	year      = {2010},
	publisher = {Association for Computing Machinery},
	address   = {New York, NY, USA}
}

@article{wang1996,
	title     = {Beyond {{Accuracy}}: {{What Data Quality Means}} to {{Data Consumers}}},
	shorttitle = {Beyond {{Accuracy}}},
	author    = {Wang, Richard Y. and Strong, Diane M.},
	year      = 1996,
	month     = mar,
	journal   = {Journal of Management Information Systems},
	volume    = {12},
	number    = {4},
	pages     = {5--33},
	issn      = {0742-1222, 1557-928X},
	doi       = {10.1080/07421222.1996.11518099},
	urldate   = {2024-12-13},
	langid    = {english}
}

@article{wand1996,
	title     = {Anchoring Data Quality Dimensions in Ontological Foundations},
	author    = {Wand, Yair and Wang, Richard Y.},
	year      = 1996,
	month     = nov,
	journal   = {Communications of the ACM},
	volume    = {39},
	number    = {11},
	pages     = {86--95},
	issn      = {0001-0782, 1557-7317},
	doi       = {10.1145/240455.240479},
	urldate   = {2024-08-28},
	langid    = {english}
}

@article{strong1997,
	author    = {Strong, Diane M. and Lee, Yang W. and Wang, Richard Y.},
	title     = {Data quality in context},
	year      = {1997},
	issue_date = {May 1997},
	publisher = {Association for Computing Machinery},
	address   = {New York, NY, USA},
	volume    = {40},
	number    = {5},
	issn      = {0001-0782},
	url       = {https://doi.org/10.1145/253769.253804},
	doi       = {10.1145/253769.253804},
	journal   = {Commun. ACM},
	month     = may,
	pages     = {103–110},
	numpages  = {8}
}

@article{batini2009,
	title     = {Methodologies for Data Quality Assessment and Improvement},
	author    = {Batini, Carlo and Cappiello, Cinzia and Francalanci, Chiara and Maurino, Andrea},
	year      = 2009,
	month     = jul,
	journal   = {ACM Computing Surveys},
	volume    = {41},
	number    = {3},
	pages     = {1--52},
	issn      = {0360-0300, 1557-7341},
	doi       = {10.1145/1541880.1541883},
	urldate   = {2024-12-13},
	langid    = {english}
}

@article{cai2015,
	title     = {The challenges of data quality and data quality assessment in the big data era},
	author    = {Cai, Li and Zhu, Yangyong},
	journal   = {Data science journal},
	volume    = {14},
	pages     = {2--2},
	year      = {2015}
}

@techreport{mqaf1,
	title     = {A metadata quality assurance framework},
	author    = {Kir{\'a}ly, P{\'e}ter},
	institution = {Gesellschaft f{\"u}r wissenschaftliche Datenverarbeitung mbH G{\"o}ttingen (GWDG): G{\"o}ttingen, Germany},
	address   = {G{\"o}ttingen, Germany},
	year      = {2015},
	url       = {https://pkiraly.github.io/metadata-quality-project-plan.pdf}
}

@misc{qualitysurvey,
	title     = {How to Define the Quality of Data? A Feature-Based Literature Survey}, 
	author    = {Markus Matoni and Arno Kesper and Gabriele Taentzer},
	year      = {2025},
	url       = {https://arxiv.org/abs/2504.01491}, 
}

@misc{qualitydimensions,
	title     = {Quality of Descriptive Information on Cultural Heritage Objects: Definition and Empirical Evaluation}, 
	author    = {Markus Matoni and Arno Kesper and Gabriele Taentzer},
	year      = {2026},
	url       = {https://arxiv.org/abs/2602.21249}, 
}

\end{document}